# An Innovative Approach to Balancing Chemical-Reaction Equations: A Simplified Matrix-Inversion Technique for Determining The Matrix Null Space

**Lawrence R. Thorne**

*Sandia National Laboratories, Livermore, CA 94550, lrthorn@sandia.gov*

**Abstract:** I propose a novel approach to balancing equations that is applicable to all chemical-reaction equations; it is readily accessible to students via scientific calculators and basic computer spreadsheets that have a matrix-inversion application. The new approach utilizes the familiar matrix-inversion operation in an unfamiliar and innovative way; its purpose is not to identify undetermined coefficients as usual, but, instead, to compute a matrix null space (or matrix kernel). The null space then provides the coefficients that balance the equation. Indeed, the null space contains everything there is to know about balancing any chemical-reaction equation!

## Introduction

Traditional matrix-inversion methodologies for balancing chemical-reaction equations can be used to balance only a restricted class of equations (i.e., those which have a single, unique solution), and many reaction equations do not fall into this category. The new matrix null-space method is applicable to all chemical-reaction equations without exception. It is adept even at notoriously difficult categories of reactions such as those which involve large numbers of complex molecules and those which require balancing both mass and electrical charge simultaneously. The new method also determines the number of balanced solutions (i.e., 0, 1 or an infinite number) applicable to a particular equation even before the equation's balancing coefficients are computed; it can do so because it determines the number of possible solutions from the dimensionality of the null space.

Unfortunately, the capability to compute the null space is usually found only in specialized software applications. By contrast, the capability to perform standard matrix inversion is a widely available, built-in function of scientific calculators and basic spreadsheets. So using common matrix inversion to compute the null space makes a hugely sophisticated technique available through the simplest of means!

Two of the first concepts that chemistry students encounter are the chemical-reaction equation and the law of combining, i.e., conservation of mass applied to chemical reactions [1]. Chemical-reaction equations are necessary for calculating the proportions and quantities of products produced and reactants consumed if a chemical reaction proceeds (as permitted by thermodynamics and kinetics) to completion or equilibrium. The law of combining defines the mathematics of such chemical reactions and states that the atoms, molecules, electrons and ions participating in a chemical reaction must combine in ratios of whole numbers (integers) to produce products.

A popular, casual approach to coping with the challenges of chemical-reaction equations is "balancing by inspection" [2, 3]. The standard balancing-by-inspection approach is to make successive, hopefully intelligent guesses at the coefficients that will balance an equation, continuing until balance is achieved. This can be a straightforward, speedy approach for simple equations. But it rapidly becomes both laborious and haphazard for complex reactions that involve many reactants and products or that require balancing both mass and charge simultaneously, such as oxidation-reduction equations. Admittedly, more systematic approaches to implementing the balancing-by-inspection method have been published for use with such challenging equations, and these are typically slightly faster and more reliable than simple inspection/guesswork. They usually recommend identifying either the "least adjustable" coefficients or the coefficients for a set of sub-reactions and balancing them by inspection first, then balancing the undetermined coefficients, algebraically, in a final step. But even these improved approaches overlook a critical problem. Balancing by inspection does not produce a systematic evaluation of all of the sets of coefficients that would potentially balance an equation; in other words, the technique encourages the notion that there is one, and only one, correct solution for any skeletal equation. In fact, there may be no possible solution, one unique solution, or an infinite number of solutions.

Another common method of balancing chemical-reaction equations is the algebraic approach [4, 5]. In this approach, coefficients are treated as unknown variables or undetermined coefficients whose values are found by solving a set of simultaneous equations [6, 7]. And a routine way of solving simultaneous equations is to employ matrix inversion. Overall, this algebraic approach/matrix-inversion methodology presents far fewer opportunities to make arithmatic errors than does the balancing-by-inspection approach, and it can handle many complex reactions easily. In addition, the matrix-inverse function is familiar to many chemistry students; and it is widely available on hand-held scientific calculators and computer spreadsheet programs, which makes it computationally convenient. Unfortunately, however, like the balancing-by-inspection approach, the algebraic approach using matrix inversion for balancing chemical equations is not general; it can be used to balance only a restricted class of reactions—specifically, those for which the number of simultaneous equations (derived by mass and charge balance) required for a solution is one less than the number of balancing coefficients in the original chemical equation. In other words, the approach works only for equations which have a single, unique solution. And in reality, many, important chemical-reaction equations do not fall into this category.

An alternative, powerful matrix method--the matrix null-space approach--is general, but it isn't simple. And since it isn't a standard function/application for calculators and spreadsheets, it's used almost exclusively by professionals. But if the common matrix-inversion operation could be used to find the uncommon matrix null space, even non-professionals could balance all chemical-reaction equations easily…and now they can.

**Discussion**

The matrix inversion method introduced here determines the matrix null-space in a clever way. Instead of inverting a matrix to solve simultaneous equations, the new methodology inverts an innovative, augmented, row-echelon matrix to compute the null space of the original chemical-composition matrix! The null space then provides the coefficients that balance the chemical equation…and a wealth of additional information as well [8–10]!

**Basic Steps in Balancing A Chemical-Reaction Equation With The Simplified Matrix Null-Space Method**

(a) Construct a chemical-composition matrix for the chemical-reaction equation.
(b) Determine the nullity, or dimensionality, of the matrix null space of the chemical- composition matrix.
(c) Augment the matrix with a number of rows equal to the nullity number.
(d) Compute the matrix inverse of the augmented matrix by using the built-in functions of a scientific calculator or computer spreadsheet program.
(e) Extract the null-space basis vectors from the inverted matrix. (The vectors will be the columns at the far right of the inverted matrix; the number of columns included—0, 1 or more—should equal the nullity of the row-echelon matrix.) *This defines the null space of the original chemical-composition matrix!*
(f) Take the transpose of the null space vectors. (Each transposed null-space vector is proportional to a set of coefficients which will ultimately balance the chemical-reaction equation.)
(g) Make the smallest number in each transposed vector equal to 1 by dividing each vector element by the element of the smallest magnitude. (These scaled vectors are the coefficients that balance the equation, term by term, left to right.)
(h) Construct a new chemical-reaction equation by placing positive terms on one side, negative terms on the other. *The equation is now balanced!*

To illustrate these eight, basic steps, an example of balancing a chemical-reaction equation using the matrix null-space method follows. Consider, for instance, the following chemical-reaction equation:

$$KI + KClO_3 + HCl = I_2 + H_2O + KCl \quad (1)$$

Step #1. A chemical-composition table is a fast, straightforward way to construct a chemical-composition matrix by organizing the elemental constituents of an equation of interest. A table for Eq. 1 looks like this:

|    | KI | KClO$_3$ | HCl | I$_2$ | H$_2$O | KCl |
|----|----|----------|-----|-------|--------|-----|
| K  | 1  | 1        | 0   | 0     | 0      | 1   |
| I  | 1  | 0        | 0   | 2     | 0      | 0   |
| O  | 0  | 3        | 0   | 0     | 1      | 0   |
| H  | 0  | 0        | 1   | 0     | 2      | 0   |
| Cl | 0  | 1        | 1   | 0     | 0      | 1   |

Entries from this table then become the values of the chemical-composition matrix, as shown below.

$$\begin{bmatrix} 1 & 1 & 0 & 0 & 0 & 1 \\ 1 & 0 & 0 & 2 & 0 & 0 \\ 0 & 3 & 0 & 0 & 1 & 0 \\ 0 & 0 & 1 & 0 & 2 & 0 \\ 0 & 1 & 1 & 0 & 0 & 1 \end{bmatrix}$$

The Chemical-Composition Matrix
For Equation/Reaction (1)

A chemical-composition matrix specifies the numbers of atoms of each chemical element which make up each of the reactants and products specified in a given reaction equation. The matrix null-space method does not require distinguishing between reactants and products in the composition matrix, so all matrix elements are simply specified either as zero or as positive numbers, which need not be integers but usually are.

Step #2. The nullity, or dimensionality, of the null space of the chemical-composition matrix is needed to construct the partitioned matrix detailed in Step #3 below. The nullity can be determined by taking the difference between the number of chemical species in, and the rank of, the chemical-composition matrix. The number of chemical species can be determined simply by counting the number of columns in the matrix. The rank can be determined in at least two (2) ways. The faster, easier approach is to employ the common spreadsheet function MDETERM to find the number of rows or columns in the largest, square sub-matrix (of the chemical-composition matrix) that has a non-zero determinant. The alternative, slower, manual approach is outlined in the Appendix hereto.

Note: for the chemical-composition matrix shown above, the number of chemical species is 6; the rank is 5; and the nullity, therefore, is 1. This is true because the largest, square, non-zero-determinant sub-matrix is a 5 × 5 matrix.

Step #3. The next step is to augment the bottom of the initial chemical-composition matrix with a number of rows equal to the nullity number. The added rows should form a partitioned matrix composed of a rightmost, square, identity sub-matrix (with dimensionality of one or higher) and a leftmost, rectangular sub-matrix that contains only zeroes. After this augmentation, the resultant matrix will be square. For Eq. 1, the appropriately-augmented matrix is:

$$\begin{bmatrix} 1 & 1 & 0 & 0 & 0 & 1 \\ 1 & 0 & 0 & 2 & 0 & 0 \\ 0 & 3 & 0 & 0 & 1 & 0 \\ 0 & 0 & 1 & 0 & 2 & 0 \\ 0 & 1 & 1 & 0 & 0 & 1 \\ 0 & 0 & 0 & 0 & 0 & 1 \end{bmatrix}$$

Augmented Chemical Composition Matrix

Note: This augmentation step is inapplicable if the chemical-composition matrix is square to begin with, as it would be, for instance, for ion exchange reactions. In that event, it is necessary to reduce the matrix to a row-echelon form; for details on this procedure, please see the Appendix. The row-echelon form of the chemical-composition matrix will have one or more bottom rows containing only zeroes; to augment this form correctly, the zero rows must be replaced by a partitioned matrix of the proper dimension, as described above.

Step #4. *At this point, the augmented matrix is used as input for a typical scientific calculator or computer spreadsheet, and the machine calculates the matrix inverse, which is the hard part of this process!* (Note: A glutton for punishment can do this step by hand using Cramer's Rule.)

$$\begin{bmatrix} 1 & 0 & -0.286 & 0.143 & -0.143 & -0.857 \\ 0 & 0 & 0.286 & -0.143 & 0.143 & -0.143 \\ 0 & 0 & -0.286 & 0.143 & 0.857 & -0.857 \\ -0.5 & 0.5 & 0.143 & -0.071 & 0.071 & 0.429 \\ 0 & 0 & 0.143 & 0.429 & -0.429 & 0.429 \\ 0 & 0 & 0 & 0 & 0 & 1 \end{bmatrix}$$

Inverse of the Augmented Chemical-Composition Matrix

Step #5. The important null-space vector(s) are now revealed as the rightmost column(s) of the inverted matrix; the number of vectors should equal the nullity of the original chemical-composition matrix--in this example, 1 column/vector.

$$\begin{bmatrix} 1 & 0 & -0.286 & 0.143 & -0.143 & \boxed{-0.857} \\ 0 & 0 & 0.286 & -0.143 & 0.143 & \boxed{-0.143} \\ 0 & 0 & -0.286 & 0.143 & 0.857 & \boxed{-0.857} \\ -0.5 & 0.5 & 0.143 & -0.071 & 0.071 & \boxed{0.429} \\ 0 & 0 & 0.143 & 0.429 & -0.429 & \boxed{0.429} \\ 0 & 0 & 0 & 0 & 0 & \boxed{1} \end{bmatrix}$$

Highlighted Null-Space Vector

Step #6. The transpose of the null-space vector(s) is taken.

$$[-0.857 \quad -0.143 \quad -0.857 \quad 0.429 \quad 0.429 \quad 1]$$

Transpose of the Null-Space Vector

Step #7. Then each vector element is scaled (or divided) by the element of the smallest magnitude (0.167 in the example) so that the smallest number in each transposed vector is equal to 1.

$$[-6 \quad -1 \quad -6 \quad 3 \quad 3 \quad 7]$$

Scaled, Transpose of the Null-Space Vector

The scaled vector(s) yields the coefficients that balance the equation, term by term, left to right, as they appear at the top of the chemical-composition table.

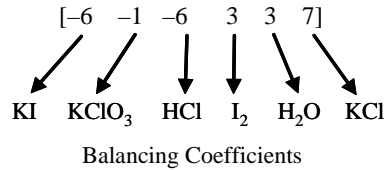

Balancing Coefficients

Step #8. And finally, the new equation is written in standard form by placing positive terms (i.e., of the scaled, transposed, null-space vector elements) on one side, negative terms on the other. *And the equation is balanced!*

$$\underline{6}\ KI + \underline{1}\ KClO_3 + \underline{6}\ HCl = \underline{3}\ I_2 + \underline{3}\ H_2O + \underline{7}\ KCl$$

Some skeletal chemical-reaction equations can be complicated to deal with because they can be balanced by an infinite number of sets of coefficients. The example below is one such equation (6):

$$CO + CO_2 + H_2 = CH_4 + H_2O \qquad (2)$$

The new, null-space method proposed here finds two, independent sets of balancing coefficients for this equation.

$$CO_2 + CH_4 = 2CO + 2H_2 \qquad (2a)$$

$$CO + H_2O = CO_2 + H_2 \qquad (2b)$$

Linear combinations of these two sets of balancing coefficients also balance the equation, and, therefore, there exists an infinite number of balancing sets. For example, the combination formed by 2 times Eq. (2a) plus 3 times Eq. (2b) balances the skeletal equation.

$$CO + CO_2 + 7H_2 = 2CH_4 + 3H_2O \qquad (2c)$$

Eq. 2c contains all of the chemical species of the original, skeletal equation; whereas the two, independent equations, 2a and 2b, do not. $H_2O$ does not appear in Eq. 2a, and $CH_4$ does not appear in Eq. 2b.

Rarely, one encounters a skeletal chemical equation that cannot be balanced, such as the one below. However, this type of equation can be identified easily because its nullity will be zero. This condition will be manifest if: (1) the determinant of the chemical-composition matrix is not equal to zero; or, (2) there are no rows comprised entirely of zeroes in the row-echelon form of the chemical-composition matrix.

$$FeS_2 + HNO_3 = Fe_2(SO_4)_3 + NO + H_2SO_4 \qquad (3)$$

$$\begin{bmatrix} 1 & 0 & 0 & 0 & 0 \\ 0 & 1 & 0 & 0 & 0 \\ 0 & 0 & 1 & 0 & 0 \\ 0 & 0 & 0 & 1 & 0 \\ 0 & 0 & 0 & 0 & 1 \end{bmatrix}$$

### The Row-Echelon Matrix

Another quick way to see if a chemical-reaction equation can be balanced is to try to compute its inverse. If a chemical-composition matrix is square and can be inverted, the chemical-reaction equation from which the composition matrix was derived <u>cannot</u> be balanced unless more species are added to the reaction. In Eq. 3 above, for instance, the addition of water would permit the equation to be balanced, which would be a reasonable adjustment if the reaction took place in an aqueous solution.

Numerous and complex reactants and products can increase the challenge of balancing a chemical-reaction equation dramatically. And the presence of electrical charge can introduce even further complexity, since charge and mass must be balanced simultaneously. Eq. 4 below is a diabolical example (11) of just such an equation. But the new matrix null-space method (developed, in this example, with the row-echelon approach) copes with it easily!

$$Cr_7N_{66}H_{96}C_{42}O_{24} + MnO_4^- + H^+ =$$

$$Cr_2O_7^{2-} + Mn^{2+} + CO_2 + NO_3^- + H_2O \qquad (4)$$

The chemical-composition table for Eq. 4 follows. Note: electrical charge is treated as an ordinary element, although its positive or negative sign must be specified.

|  | $Cr_7N_{66}H_{96}C_{42}O_{24}$ | $MnO_4^-$ | $H^+$ | $Cr_2O_7^{2-}$ | $Mn^{2+}$ | $CO_2$ | $NO_3^-$ | $H_2O$ |
|---|---|---|---|---|---|---|---|---|
| Cr | 7 | 0 | 0 | 2 | 0 | 0 | 0 | 0 |
| N | 66 | 0 | 0 | 0 | 0 | 0 | 1 | 0 |
| H | 96 | 0 | 1 | 0 | 0 | 0 | 0 | 2 |
| C | 42 | 0 | 0 | 0 | 0 | 1 | 0 | 0 |
| O | 24 | 4 | 0 | 7 | 0 | 2 | 3 | 1 |
| Mn | 0 | 1 | 0 | 0 | 1 | 0 | 0 | 0 |
| Charge | 0 | -1 | 1 | -2 | 2 | 0 | -1 | 0 |

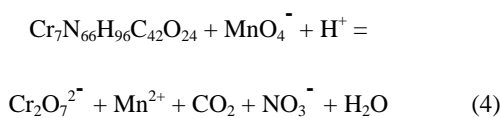

The Chemical-Composition Matrix          The Row-Echelon Matrix

The Augmented Row-Echelon Matrix     Inverse of the Row-Echelon Matrix

Note: The rank of the row-echelon matrix is 7, since further reduction by elementary row operations does not produce any rows of exclusively zeroes; its nullity, therefore, is equal to 1.

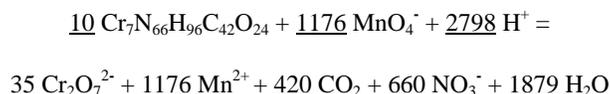

Highlighted Null-Space Vector

And after dividing the null-space vector by the element of the smallest magnitude, the vector is multiplied by 10 to obtain the whole numbers of the scaled, transposed row vector.

[–10   –1176   –2798   35   1176   420   660   1879]

Scaled Transpose Of The Null-Space Vector

Therefore:

<u>10</u> $Cr_7N_{66}H_{96}C_{42}O_{24}$ + <u>1176</u> $MnO_4^-$ + <u>2798</u> $H^+$ =

<u>35</u> $Cr_2O_7^{2-}$ + <u>1176</u> $Mn^{2+}$ + <u>420</u> $CO_2$ + <u>660</u> $NO_3^-$ + <u>1879</u> $H_2O$

And the equation is balanced!

For the sake of thoroughness, it should be noted that there are, of course, other ways to identify the matrix null space. These include singular value decomposition; eigenvalue decomposition; the generalized matrix inverse (the Moore-Penrose Pseudo Inverse); *LU* decomposition; and *QR* decomposition. These are legitimate routes to the null space. But they are all much more complex and time-consuming than the short-cut approach proposed herein. Furthermore, they can't be computed on simple calculators and basic spreadsheets.

### Conclusion

This new, simplified, matrix null-space method of balancing chemical-reaction equations is an important advance in terms of universal applicability, fast computation and high accuracy. Furthermore, in this era of unprecedented student access to digital compute power, it's straightforward to employ! And it's fun!

**Acknowledgment.** Portions of this research were supported by Sandia. Sandia is a multiprogram laboratory operated by Sandia Corporation, a lockheed Martin Company, for the United States Department of Energy's National Nuclear Security Administration under Contract DE-AC04-94AL85000.

**Supporting Materials.** An Appendix containing A Manual Approach To Determining The Rank Of A Chemical-Composition Matrix is available for download at (http://dx.doi.org/10.1333/s00897102277a).

# Supporting Material/Appendix

A Manual Approach To Determining The Rank Of A
Chemical-Composition Matrix
Lawrence R. Thorne

A good way to determine the rank of a chemical-composition matrix is to reduce the composition matrix to a row-echelon form, then to determine its nullity (i.e., the dimensionality of the matrix null space). This approach, as detailed below, is referred to as Gaussian Elimination.

The chemical-composition matrix is reduced to row-echelon form by using elementary row operations on the matrix. The matrix need not be reduced completely to its unique, reduced row-echelon form; any of the many standard row-echelon forms is adequate.

A matrix is in row-echelon form if all non-zero rows are above any rows of all zeroes and if the leftmost element of a row is always to the right of the leftmost element of the row above it. The operations used to transform a matrix into row-echelon form using Gaussian Elimination consist of the elementary row operations:

A.  Interchanging any two rows;
B.  Multiplying all elements of a row by the same number;
C.  Replacing a row by the sum of that row and a multiple of another row.

The order in which the various row operations are performed does not matter. However, a convenient way to proceed is to:

1.  Select a row that has a non-zero leftmost element and more zero elements than the other rows in the matrix.
2.  Use this row to make the leftmost element of all the other rows equal zero by sequential use of row operation (C) above.
3.  Repeat steps (1) and (2) for the elements that are one element to the right of the leftmost elements.
4.  Then repeat (1) and (2) for the elements two elements to the right, and so on, until no further zeros can be produced in any columns.
5.  Produce the final echelon structure by using row interchanges. The unique, *row-reduced* echelon form is produced by using the elementary row operation (B) to make the leftmost, non-zero element in each row equal to 1.

$$\begin{bmatrix} 1 & 1 & 0 & 0 & 0 & 1 \\ 1 & 0 & 0 & 2 & 0 & 0 \\ 0 & 3 & 0 & 0 & 1 & 0 \\ 0 & 0 & 1 & 0 & 2 & 0 \\ 0 & 1 & 1 & 0 & 0 & 1 \end{bmatrix} \xrightarrow[\text{rows 2 \& 3}]{\text{a. Interchange}} \begin{bmatrix} 1 & 1 & 0 & 0 & 0 & 1 \\ 0 & 3 & 0 & 0 & 1 & 0 \\ 1 & 0 & 0 & 2 & 0 & 0 \\ 0 & 0 & 1 & 0 & 2 & 0 \\ 0 & 1 & 1 & 0 & 0 & 1 \end{bmatrix} \xrightarrow[\text{rows 3 \& 4}]{\text{b. Interchange}} \begin{bmatrix} 1 & 1 & 0 & 0 & 0 & 1 \\ 0 & 3 & 0 & 0 & 1 & 0 \\ 0 & 0 & 1 & 0 & 2 & 0 \\ 1 & 0 & 0 & 2 & 0 & 0 \\ 0 & 1 & 1 & 0 & 0 & 1 \end{bmatrix}$$

Chemical-Composition Matrix

c. Multiply row 2 by 1/3
d. Multiply row 2 by -1 & add to row 1
e. Multiply row 1 by -1 & add to row 4
f. Multiply row 4 by ½
g. Multiply row 2 by -1 & add to row 5
h. Multiply row 3 by -1 & add to row 5
i. Multiply row 5 by 3/7
j. Multiply row 5 by -6 & add to row 4
k. Multiply row 5 by -2 & add to row 3
l. Multiply row 5 by 1/3 & add to row 2
m. Multiply row 5 by 1/3 & add to row 1

$$\longrightarrow \begin{bmatrix} 1 & 0 & 0 & 0 & 0 & 6/7 \\ 0 & 1 & 0 & 0 & 0 & 1/7 \\ 0 & 0 & 1 & 0 & 0 & 6/7 \\ 0 & 0 & 0 & 1 & 0 & -3/7 \\ 0 & 0 & 0 & 0 & 1 & -3/7 \end{bmatrix}$$

Reduced Row-Echelon Matrix

Transformation Of The Chemical-Composition Matrix, For Eq. (1), To Its Reduced Row-Echelon Form

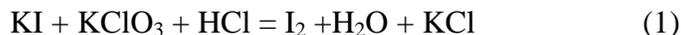

$$KI + KClO_3 + HCl = I_2 + H_2O + KCl \qquad (1)$$

The row-echelon form of the matrix yields critical information. It reveals both the rank and the nullity of the chemical-composition matrix. The rank equals the number of rows that contain numbers other than zero, (rank = 5 in the example above); it also determines how the row-echelon matrix can be augmented, as it must be, for use in the new null-space approach. The nullity equals the number of either the rows or columns in the matrix, whichever total is greater, minus the rank, (nullity equals 6 -5 = 1 in the example above). In other words, the nullity is the column dimension of the null space. Importantly, the nullity of the matrix discloses the number of possible solutions that exist for—or how many sets of coefficients would balance—the chemical-reaction equation of interest: none, one or an infinite number. When the nullity is zero, there are no solutions that will balance the equation. When it's one, there is a single (unique) solution. When the nullity is two (2) or more, there are an infinite number of solutions. (Note: the concept of many, correct answers to a single mathematical problem is an odd one to consider. For purposes of this discussion, i.e., when balancing chemical-reaction equations, one need only understand that if two, independent sets of coefficients will balance an equation, then all linear combinations of those sets will also balance the same equation--an infinite number of solutions.)